\documentclass[aps,twocolumn]{revtex4}%
\usepackage{amsfonts}
\usepackage{amsmath}
\usepackage{amssymb}
\usepackage{version}
\usepackage{graphicx}%
\includeversion{New_connection}
\excludeversion{Old_connection}

\begin{document}
\preprint{HEP/123-qed}
\title[Non-quantum escape model]{A classical statistical model for distributions of escape events in swept-bias
Josephson junctions}
\author{James A. Blackburn}
\affiliation{Physics \& Computer Science, Wilfrid Laurier University, Waterloo, Ontario, Canada}
\author{Matteo Cirillo}
\affiliation{Dipartimento di Fisica and MINAS-Lab, Universit\`{a} di Roma \textquotedblleft
Tor Vergata\textquotedblright\ I-00133 Roma, Italy}
\author{Niels Gr\o nbech-Jensen}
\affiliation{Department of Applied Science, University of California, Davis, California 95616}
\affiliation{The Niels Bohr International Academy, The Niels Bohr Institute, Blegdamsvej
17, 2100 Copenhagen, Denmark}

\begin{abstract}
We have developed a model for experiments in which the bias current applied to
a Josephson junction is slowly increased from zero until the junction switches
from its superconducting zero-voltage state, and the bias value at which this
occurs is recorded. Repetition of such measurements yields experimentally
determined probability distributions for the bias current at the moment of
escape. Our model provides an explanation for available data on the
temperature dependence of these escape peaks. When applied microwaves are
included we observe an additional peak in the escape distributions and
demonstrate that this peak matches experimental observations. The results
suggest that experimentally observed switching distributions, with and without
applied microwaves, can be understood within classical mechanics and may not
exhibit phenomena that demand an exclusively quantum mechanical interpretation.

\end{abstract}
\maketitle

\section{Introduction}

In 1974, Fulton \& Dunkleberger \cite{Fulton} demonstrated the way in which a
biased Josephson junction could be thermally excited from its zero voltage
state. \ More precisely, they conducted experiments in which the bias current
was steadily increased from zero towards its critical current. \ In the
absense of any noise, thermal or otherwise, the junction would not switch
until the bias current reached the critical value. However, with thermal
noise, junctions were observed to switch with high probability at bias
currents that were slightly less than the critical value. \ This type of
experiment has proven to be an extremely useful tool for probing the details
of effective models of the junctions themselves. \ Later work by Voss \& Webb
\cite{Voss} extended the experiments to much lower temperatures and found what
was interpreted to be evidence that the junction had entered a mode where an
escape might be treated as Macroscopic Quantum Tunneling \ (MQT) out of the
effective potential well associated with junction phase dynamics.

Once the idea of macroscopic quantum behavior for Josephson junctions at low
temperatures became accepted, the possibility of the manifestation of discrete
quantum levels within the effective potential wells received attention. The
first experiment to consider this propostion was reported by Martinis, Devoret
and Clarke (MDC) \cite{MDC}. \ In that experiment, microwaves were directed
onto a junction and the bias current was swept as before. The idea was that
the microwaves would excite transitions from a lower to a higher level within
the well, and that the macroscopic quantum variable - the junction phase -
would then tunnel out of the well from that higher level, resulting in an
escape from the zero voltage state. Single and multiple peaks in the escape
distributions of two samples were reported in their experiment. These data
were interpreted as signatures of the anticipated level transitions dictated
by quantum theory.

Previoisly we analyzed the issue of the classical resonant
frequencies in wells under both harmonic and anharmonic approximations and
found \cite{Blackburn}\ that the classical theory gave results in good
agreement with the experiments of MDC, suggesting that the classical Resistive
and Capacitive Shunted Junction (RCSJ) model for a Josephson junction might
have been dismissed prematurely in favor of the macroscopic quantum
picture\cite{Leggett}. The importance of the anharmonic component to the
potential well is evident from the fact that a measurement of escape cannot be
realized in a harmonic well and, thus, one should not expect meaningful
agreement between switching experiments and classical analysis of plasma
oscillations. This notion was first presented in Ref.~\cite{Jensen0}, where
the anharmonic theory successfully compared to accompanying experiments on
both direct and harmonic resonant switching. It should be mentioned that the
anharmonic classical RCSJ approach has since also produced good agreement with
other experimentally observed features, such as Rabi oscillations and Ramsey
fringes \cite{NGJ,Marchese} as well as tomographic reconstruction of
anticipated density matrices for a pair of capacitively coupled Josephson
junctions \cite{Blackburn2}, which originally had been interpreted exclusively
in terms of quantum entanglement.

While the non-quantized RCSJ model has proven to replicate the primary
resonant features of the experiments, available experimental reports on
switching during bias-sweep contain details not yet directly analyzed using
the non-quantized approach. One is the saturation of the width of the
switching distribution as the thermodynamic temperature is lowered,
interpreted as a signature of a quantum crossover temperature \cite{Voss};
another is the set of multiple of resonant switching peaks, interpreted as
signatures of quantized energy transitions in the potential well \cite{MDC}.

In this paper, we reconsider the evidence for the macroscopic quantum
tunneling interpretation of these experiments. \ In particular, in the spirit
of Kramers' statistical analysis \cite{Kramers}, we develop a simple model of
the swept-bias type of experiment based on classical thermal activation, and
show that it gives an excellent accounting of the peaks observed in a number
of key experiments.

\section{Modeling a Josephson Swept-Bias Experiment}

We consider Josephson junctions to be characterized by a supercurrent
$I_{C}\sin\varphi$ with a critical current $I_{C}$ and junction phase
$\varphi$, and a capacitance $C$. The associated junction plasma frequency is
$\omega_{J}=\sqrt{2\pi I_{C}/\Phi_{0}C}$, where $\Phi_{0}=h/2e$ is the flux
quantum. We assume junctions whose physical dimensions are much smaller than
the Josephson penetration depth.

From this perspective, the phase dynamics of a single Josephson junction
subjected to a dc bias is equivalent to the motion of a particle on a
washboard potential \cite{vanDuzer}. The particle sits in a potential well
which becomes shallower at high bias currents. This in turn leads to an
enhanced probability of noise activated escape. If the bias current is swept
from $0$ towards $I_{C}$, then at some moment the junction will be observed to
escape from its potential well and switch to a running state with finite
voltage. In the experiments to which we will refer, this process was repeated
many times in order to acquire a statistical profile of the distribution of
bias currents for which the escapes from the zero-voltage state occur.

For our numerical simulations, we imagine an equivalent scenario. \ Suppose
there is an ensemble of $M$ \ Josephson junctions. The bias on all junctions
starts at $0$ and is incremented in $N$ steps, with each step of duration
$\Delta t=(Nf_{S})^{-1}$ where $f_{S}$ is the sweep frequency. Each step is
assigned a channel and the total counts in that channel indicate how many
junctions have switched to a finite voltage state (escape from the potential
well) during that interval. As the bias sweep proceeds, the original ensemble
will have lost $e_{1}$ junctions in the first interval, $e_{2}$ junctions in
the second interval, and so forth. \ Consequently, at the beginning of the
$n^{th}$ bias interval, there will be $M-%
{\displaystyle\sum\limits_{j=1}^{n-1}}
e_{j}$ junctions not yet escaped. The number from this remaining pool of
junctions that will escape during the next $\Delta t$ seconds will be%
\begin{equation}
e_{n}=\left[  M-%
{\displaystyle\sum\limits_{j=1}^{n-1}}
e_{j}\right]  \ \Gamma(t_{n})\Delta t,\ n=2,3,...N \label{Eq.1}%
\end{equation}
where $\Gamma(t_{n})$ is the probability of escape per unit time in the
$n^{th}$ interval, \ Of course, the initial interval just satisfies%
\begin{equation}
e_{1}=M\Gamma(t_{1})\Delta t \label{Eq.2}%
\end{equation}
Equations (\ref{Eq.1},\ref{Eq.2}) will mimic a swept-bias experiment provided
a suitable expression is available for the escape rate $\Gamma$.

In Kramers' theory \cite{Kramers}, the thermal escape rate can be expressed%
\begin{equation}
\Gamma(t_{n})=f_{n}\exp\left(  -\frac{\Delta U_{n}}{k_{B}T}\right)
\label{Kramers}%
\end{equation}
where $f_{n}$ is the plasma frequency for the well specific to the $n^{th}$
bias interval and $\Delta U_{n}/k_{B}T$ \ is the height of the potential
barrier divided by the mean thermal energy. Voss and Webb \cite{Voss} assumed
an escape rate in this form.

However, there has been an ongoing discussion regarding the suitability of the
Kramers expression, and as Devoret, Esteve, Martinis, Cleland, and Clarke
\cite{Devoret}, and others, have pointed out, a better equation for the escape
rate, from B\"{u}ttiker, Harris, and Landauer (BHL) \cite{BHL}, is%
\begin{equation}
\Gamma_{BHL}(t_{n})=a_{t}f_{n}\exp\left(  -\frac{\Delta U_{n}}{kT}\right)
\label{GammaBHL}%
\end{equation}
where
\begin{equation}
a_{t}=\frac{4\alpha}{\left[  1+\sqrt{\left(  1+\frac{\alpha Qk_{B}T}{1.8\Delta
U_{n}}\right)  }\right]  ^{2}} \label{at}%
\end{equation}
According to Devoret \textit{et al.}\cite{Devoret}, $\alpha=1.4738$. In this
expression, $Q$ is a parameter that quantifies the dissipation in the
junction; the lower the dissipation, the larger the $Q$. Devoret \textit{et
al.} \cite{DMEC} noted: \textquotedblleft The prefactor depends only weakly on
$Q$\textquotedblright\ and, they consequently used an escape rate in the form
of Eq.(\ref{Kramers}). \ Similarly, Devoret \textit{et al}. in \cite{Devoret}
stated that the value of the prefactor $a_{t}$ is \textquotedblleft close to
unity\textquotedblright.

For these reasons, we proceed with our classical simulations of swept bias
experiments using expression (\ref{Kramers}) for the escape rate. \ We shall
consider the second-order effects of $Q$ in Section $VI$.

In the harmonic approximation%
\begin{equation}
f_{n}=f_{J}\,\sqrt[4]{1-\eta_{n}^{2}} \label{Harmonic}%
\end{equation}
with $\eta_{n}$ being the normalized bias current within the $n^{th}$ bias interval.

Combining these expressions, we obtain
\begin{equation}
\Gamma(t_{n})\Delta t=\left[  \left(  \frac{f_{J}}{Nf_{S}}\right)
\sqrt[4]{1-\eta_{n}^{2}}\right]  \exp\left(  -\frac{\Delta U_{n}}{k_{B}%
T}\right)  \label{Eq.4}%
\end{equation}
which is required in Eq.(\ref{Eq.1}). The prefactor before the exponential
represents the number of plasma oscillations that can fit within the time
window of the $n^{th}$ data acquisition channel - that is, the number of
attempts that will occur in that time window.

According to MQT theory at sufficiently low temperatures tunneling of the
phase variable becomes the dominant escape mechanism, in which case the escape
rate takes the form%

\begin{equation}
\Gamma_{q}=a_{q}f_{n}\exp\left[  -7.2\frac{\Delta U_{n}}{hf_{n}}\left[
1+\frac{0.87}{Q}+\cdots\right]  \right]  \label{MQT}%
\end{equation}
where%
\begin{equation}
a_{q}=\left[  120\pi\left(  \frac{7.2\Delta U_{n}}{hf_{n}}\right)  \right]  ^{%
\frac12
} \label{aq}%
\end{equation}
Clearly, in contrast to the classical expression Eq.(\ref{Kramers}), this
escape rate does not depend on temperature. Voss and Webb \cite{Voss} pointed
out that the transition from classical to quantum behavior should take place
around a "crossover temperature" satisfying
\begin{equation}
T_{cr}\approx\frac{hf}{7k_{B}} \label{Tcr}%
\end{equation}

\section{Microwaves OFF}

The height of the barrier at the $n^{th}$ step is given by the well known
expression%
\begin{equation}
\Delta U_{n}=2\frac{I_{C}\Phi_{0}}{2\pi}\left[  \sqrt{1-\eta_{n}^{2}}-\eta
_{n}\cos^{-1}\eta_{n}\right]  , \label{Barrier}%
\end{equation}
so%
\begin{equation}
\frac{\Delta U_{n}}{k_{B}T}=\frac{2}{\beta}\left[  \sqrt{1-\eta_{n}^{2}}%
-\eta_{n}\cos^{-1}\eta_{n}\right]  \label{Eq.5}%
\end{equation}
where%
\begin{equation}
\beta=\left(  \frac{2\pi k_{B}}{\Phi_{0}}\right)  \frac{T}{I_{C}} \label{Eq.6}%
\end{equation}

\subsubsection{Voss \& Webb (1981)}

As an example of an experimental simulation, parameters were chosen to be:
$N=5000$, $M=100,000$, $f_{J}=35.53\ GHz$, $f_{S}=10\ Hz$, numbers consistent
with the experiments of Voss and Webb \cite{Voss} (although they gave no value
for the number of channels in their system).%
\begin{figure}
[ptb]
\begin{center}
\includegraphics[
natheight=8.5in,
natwidth=11.0in,
height=2.8418in,
width=3.5016in
]%
{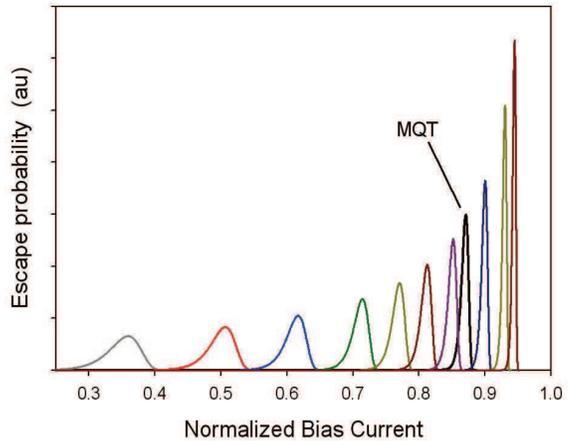}%
\caption{(color online) Simulation results for a swept-bias experiment. The
ten classical peaks had $\beta$ values (left to right): $0.0551,$ $0.0375,$
$0.0259,$ $0.0169,$ $0.0123,$ $0.0092,$ $0.00656,$ $0.00372,$ $0.00223,$
$0.00160$. The single escape peak labelled MQT was computed using expression
(\ref{MQT}) for the escape rate with $Q=50$.}%
\label{Fig.1}%
\end{center}
\end{figure}
The evolution of the peaks shown in Fig.\ref{Fig.1} is in very good agreement
with the experimental data in Fig.1 of Voss and Webb \cite{Voss}. \ Also shown
in Fig.\ref{Fig.1} is the single MQT peak from a simulation using the escape
rate expression Eq.(\ref{MQT}) with $Q=50$. \ The conclusion to be drawn is
that when the value of $\beta$ drops below a crossover equivalent, macroscopic
quantum behavior should take over from the classical escape process and the
temperature independent peak marked MQT should become frozen in place. In such
a case not only the peak widths but also the peak positions must remain
constant. Then, none of the classical peaks to the right of MQT would be
observed in an experiment.

\begin{figure}
[ptb]
\begin{center}
\includegraphics[
natheight=8.5in,
natwidth=11.0in,
height=2.6282in,
width=3.3512in
]%
{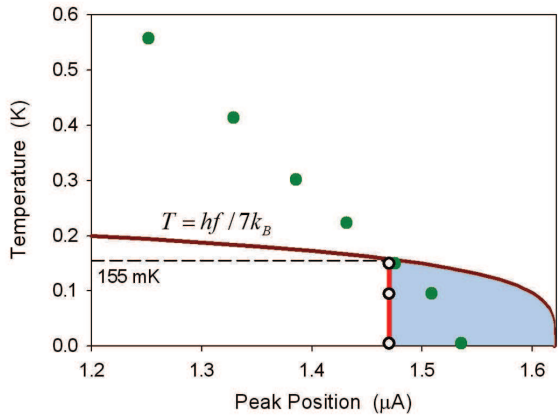}%
\caption{(color online) Dependence of the crossover temperature on bias
current, for sample parameters of Voss and Webb \cite{Voss}. \ Dots mark the
positions of the seven lowest temperature peaks shown in Fig.1 of \cite{Voss}.
At $1.470\;\mu A$ the experimental data drop below the crossover boundary.
\ According to the MQT hypothesis, the escape peaks should then become
temperature independent and peak positions would be expected to lie on the
vertical line, as depicted by open circles.}%
\label{Fig2}%
\end{center}
\end{figure}

The crossover temperature, Eq.(\ref{Tcr}), is a function of the natural
frequency of a particular well, and this in turn is controlled by the applied
bias current as specified in Eq.(\ref{Harmonic}). \ Therefore, in a swept-bias
experiment one is also sweeping the natural frequency of the continuously
varying well shape. It is simple to use Eqs. (\ref{Harmonic},\ref{Tcr}) to
plot the dependence of crossover temperature on bias current; this is shown in
Fig.\ref{Fig2}.%
The positions (bias values) of the experimental peaks, indicated by solid
dots, were manually extracted from Fig.1 in \cite{Voss} using digitizing
software \cite{note}. A vertical line marks the point at which the sample
temperature has dropped below the crossover characteristic - this occurs at
$T\approx155\;mK$ and a bias of $1.470\;\mu A$. But note that the two
experimental escape peaks for temperatures $T=95\;mK$ and $T=5\;mK$ are below
the anticipated quantum transition temperature and appear not to have frozen
at $1.470\;\mu A$, but instead continue to advance beyond the MQT stopping
point, into the shaded `forbidden' zone. \ While this progression of escape
peaks towards higher bias values is contrary to the expectations of the MQT
model, it is consistent with the classical escape model.

\begin{figure}
[ptb]
\begin{center}
\includegraphics[
natheight=8.5in,
natwidth=11.0in,
height=2.6204in,
width=3.397in
]%
{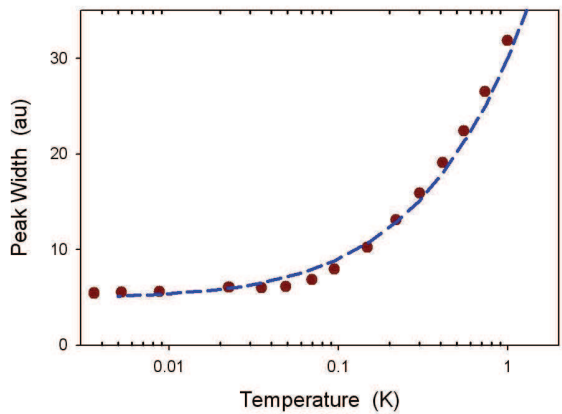}%
\caption{(color online) Dots: experimental peak widths from Fig.3 of ref.
\cite{Voss}. \ Dashed line: classical simulation with an effective sample
temperature of $T+63\;mK$ where $T$ is the bath temperature.}%
\label{Fig3}%
\end{center}
\end{figure}

The apparent saturation of the widths of the escape peaks below the crossover
tempereature, noted in \cite{Voss} was claimed to constitute the "first
compelling evidence for the existence of quantum tunneling of a macroscopic
variable". Using digitizing software, the experimental data points for the
peak widths were extracted from Fig.3 in \cite{Voss}. These points are plotted
in Fig.\ref{Fig3}.
\ More than twenty years ago, Cristiano and Silvestrini \cite{Cristiano}
proposed that the presence of some additional noise could raise the sample
temperature above the bath temperature $T$ such that $T_{eff}=T+T_{N}$. In
particular, they demonstrated that the observed temperature dependence of the
peak widths in \cite{Voss} could be replicated using $T_{N}=63\;mK$ and with
classical escape theory alone. Such an elevated sample temperature, possibly
due to self-heating, was also noted in \cite{Berkley}

We have run our simulation with $T$ replaced by $T_{eff}$ and $T_{N}=63\;mK$
and the results, virtually identical to those in \cite{Cristiano}, are shown
in Fig.\ref{Fig3}. By most standards, the agreement between experiment and
classical theory is excellent. A vote in favor of a quantum signature in these
data could only be supported by a much finer analysis including error bars in
the experimental peak widths. As it stands, the classical model fits the
experimental data exceptionally well over two orders of magnitude in bath
temperature with the above-mentioned suggested effective sample temperature.

\section{Microwaves ON}

We now consider appropriate modifications to the swept-bias model needed to
take into account the action of applied microwaves. In such experiments, a
fixed frequency microwave source is used and the bias current is ramped up.

Our numerical solutions of the equation of motion of the phase \cite{vanDuzer}
with both dc and ac bias reveal induced oscillations around the minimum in the
potential well. The amplitude of these oscillations depends on the dc bias
chosen. \ There will be a particular bias, $\eta_{res}$, at which the the
oscillation amplitude is a maximum. The question is: What is the value for
$\eta_{res}$?

In the harmonic approximation, the natural frequency of a well is related to
the dc bias through Eq.(\ref{Harmonic}). However as shown in
\cite{Blackburn,Jensen0}, when the amplitude of the phase oscillations is
large, the cubic nature of the well comes into play and an anharmonic
approximation takes the place of Eq.(\ref{Harmonic}); it is.%
\begin{equation}
f_{n}=f_{J}\sqrt{\left(  J_{0}(A)+J_{2}(A)\right)  \sqrt{1-\left(  \frac
{\eta_{n}}{J_{0}(A)}\right)  ^{2}}} \label{anharmonic}%
\end{equation}
where $J_{p}$ is the Bessel function of $p${th} order, first kind and $A$ is
the amplitude of the oscillation. Thus, the resonance frequency is depressed
for increasing oscillation amplitudes $A$. It has been found that situations,
in which resonant states produce non-zero and non-unity switching
probabilities, are given for oscillation amplitudes near the inflection point
of the potential well. In the limit $\eta\rightarrow1$, this value of $A$ is
given by the explicit expression $A^{2}\approx{\frac{4}{3}(1-\eta_{res})}$.

Setting$\ f_{n}=f_{ac}$ in Eq.(\ref{Harmonic}) would give one answer for the
resonant bias $\eta_{res}$, while Eq. (\ref{anharmonic}) would give a slightly
smaller answer. This means that without knowledge of the strength of the
microwaves at the junction, the best one can say is that $\eta_{res}$ must lie
somewhere below the value for $A=0$ (as also seen experimentally in
\cite{Jensen1}) and in the vicinity of the interval spanned by the two values
for $A=0$ and $A^{2}\approx\frac{4}{3}(1-\eta_{res})$. As an example, for a
microwave frequency $f_{ac}/f_{J}=0.350$, the limits of $\eta_{res}$ are:
$0.9925$ and $0.9887$.

Let the amplitude of the induced phase oscillation to the right of the minimum
point of any well be denoted $\delta\varphi_{n}$. For not too large
excitations, $\delta\varphi$ has a bell-shaped distribution, centered at
$\eta_{res}$. At $\eta_{res}$ the ac field is transferring a maximum amount of
energy into the junction. On either side of this optimum bias, the amplitude
of the phase oscillations diminishes and the absorbed energy declines. For a
bell shaped distribution of $\delta\varphi$ we used the following heuristic
expression,
\begin{equation}
\delta\varphi_{n}=a\frac{b^{2}}{\left(  \eta_{n}-\eta_{res}\right)  ^{2}%
+b^{2}} \label{dphi}%
\end{equation}
There are two parameters here: $b$ which sets the sharpness of the
distribution, and $a$ which sets the peak value.

Whenever sustained phase oscillations are induced, the added energy is%
\begin{align}
\frac{\Delta U_{1n}}{k_{B}T}  &  =\beta^{-1}\left\{  \left[  -\cos
\varphi_{\max}-\eta_{n}\varphi_{\max})\right]  \right. \nonumber\\
&  -\left.  \left[  -\cos\varphi_{\min}-\eta_{n}\left(  \varphi_{\min}\right)
\right]  \right\}  \label{eq.7}%
\end{align}
\allowbreak\linebreak with $\varphi_{\min}=\sin^{-1}\eta_{n}$ and
$\varphi_{\max}=\varphi_{\min}+\delta\varphi_{n}$. This will \textit{reduce}
the escape barrier in the $n^{th}$ interval to an effective value%
\begin{equation}
\frac{\Delta U_{eff}}{k_{B}T}=\frac{\Delta U_{n}}{k_{B}T}-\frac{\Delta U_{1n}%
}{k_{B}T} \label{Ueff}%
\end{equation}
and this is what thermal noise needs to overcome. \ This effective barrier
height replaces the original in Eq.(\ref{Eq.4}) and then the simulation can
proceed as before.

We now apply this simulation to several sets of published data. \ First we
consider experiments which showed only a single microwave induced escape peak.

\subsection{Single Microwave Induced Peaks}

As noted already, swept bias experiments yield histograms for the escape
probability. \ Some authors prefer to convert such data to escape rates
$\Gamma$ as a function of bias current. Then a \textit{relative rate}, with
and without microwaves, $\left[  \Gamma(P)-\Gamma(0)\right]  /\Gamma(0)$ may
be plotted. This has the effect of stripping away the thermal escape peak (as
discussed in Sections 3 \& 4), thereby isolating purely microwave induced phenomena.

\subsubsection{Martinis, Devoret \& Clarke (1985)}

\begin{figure}
[ptb]
\begin{flushleft}
\includegraphics[
natheight=11.0in,
natwidth=8.5in,
height=4.2255in,
width=3.2387in
]%
{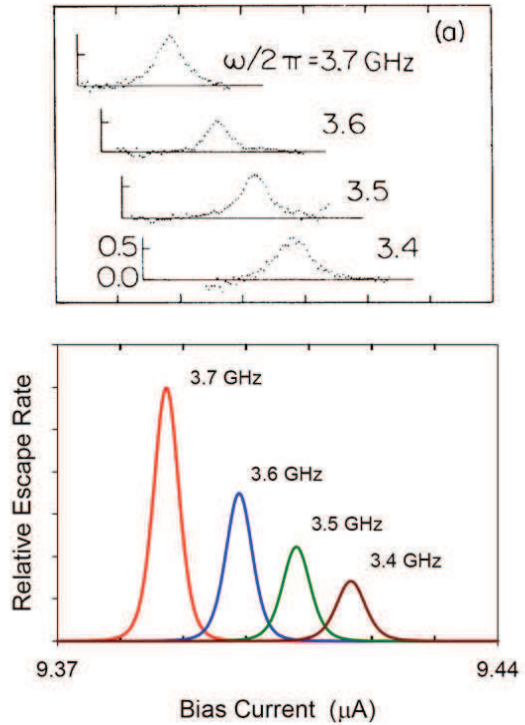}%
\caption{(color online) Comparison of the experimental results of Martinis,
Devoret, and Clarke \cite{MDC}(upper) with our simulation (lower).}%
\label{Fig4}%
\end{flushleft}
\end{figure}

Consider the results presented in Martinis, Devoret, and Clarke \cite{MDC}.
Their Fig.3 is reproduced in the upper panel of Fig.\ref{Fig4}. The junction
was characterized by the following parameter values: $I_{C}=9.489\ \mu A$ and
$C=6.35\ pF$. \ This gives a junction zero bias plasma frequency
$f_{J}=10.72\ GHz$; hence the microwave frequencies of $3.7,\ 3.6,3.5,$and
$3.4GHz$ correspond to $0.3451,0.3358,0.3265$ and $0.3172$ in dimensionless
form. This experiment was carried out at $T=18mK$.

For comparison of these experiments with classical results, we simply make use
of the escape rate expression Eq.(\ref{Kramers}) with a barrier given by
Eq.(\ref{Eq.5}) in the absense of microwaves, or with a reduced effective
barrier given by Eq.(\ref{Ueff}) when microwaves are present. For each of the
four microwave frequencies, the anharmonic result Eq.(\ref{anharmonic}) was
used to obtain the bias current $\eta_{res}$ which selects the well that is
resonant with the excitation. With this normalized bias, Eq.(\ref{eq.7})
together with Eq.(\ref{dphi}) permits the reduced barrier height to be calculated.

The remainder of the parameters chosen to match the situation were: $a=0.10$,
$b=0.0010$ and $\beta=0.0000796$. \ The results for the classical escape rate
calculations are shown in the lower panel of Fig.\ref{Fig4}
As can be seen, the classical results agree very well with the experimentally
observed peaks in the escape rates for these four microwave frequencies.

\subsubsection{Thrailkill, Lambert, Carabello \& Ramos (2009)}

\begin{figure}
[ptb]
\begin{center}
\includegraphics[
natheight=11.0in,
natwidth=8.5in,
height=4.6743in,
width=3.0355in
]%
{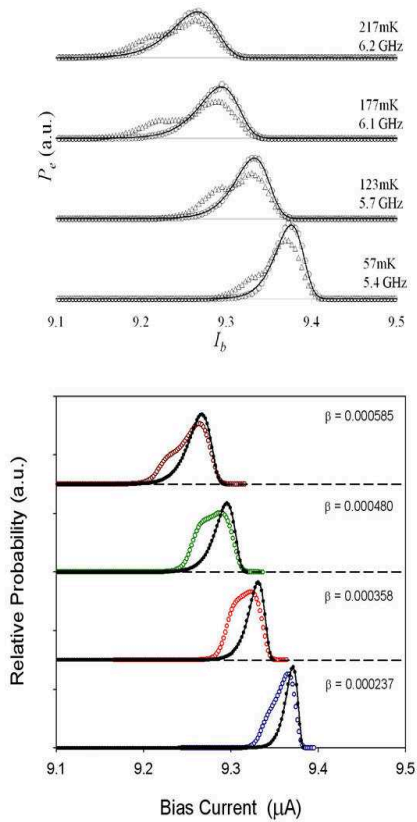}%
\caption{(color online) Comparison of experimental data from Thrailkill
\textit{et al}. \cite{Thrailkill}(upper) and our simulation results (lower)
where closed circles indicate results without microwaves.}%
\label{Fig5}%
\end{center}
\end{figure}

Thrailkill \textit{et al}. \cite{Thrailkill} carried out swept-bias
experiments on a Josephson junction characterized by $I_{C}=9.485\ \mu A$ and
$C=4.7\ pF$. The Josephson plasma frequency was thus $f_{J}=12.46\ GHz$. The
escape probability distributions were measured at several different
combinations of temperature and microwave frequency, as shown in the upper
panel of Fig.\ref{Fig5}.
For the classical simulations, the parameter values were: $a=0.09$, $b=0.003$
(equivalent to a half width in the distribution of $0.028\ \mu A$) and from
top to bottom, $\eta_{res}=0.972,0.976,0.980,0.984$. The simulation results
clearly are in very good agreement with the experimental data. Note that the
ac resonance and the thermal peak are nearly on top of each other, so the
peaks almost merge. \ Also, the simulation exhibits the same shifting effect
as in the experiments - the thermal peak without microwaves becomes displaced
slightly towards lower bias values when the microwaves are turned on.

\subsection{Multiple Microwave Induced Peaks}

Every swept bias experiment with microwaves present yields at least one ac
induced escape peak. \ This includes both Figs.2 and 3 in \cite{MDC}, Fig.3 in
\cite{Berkley}, Fig.2 in \cite{Thrailkill}, and Fig.6.5 in \cite{Dutta}. But
only in two instances was a second microwave peak in evidence.

\begin{figure}
[ptb]
\begin{center}
\includegraphics[
natheight=8.5in,
natwidth=11.0in,
height=2.3021in,
width=3.2612in
]%
{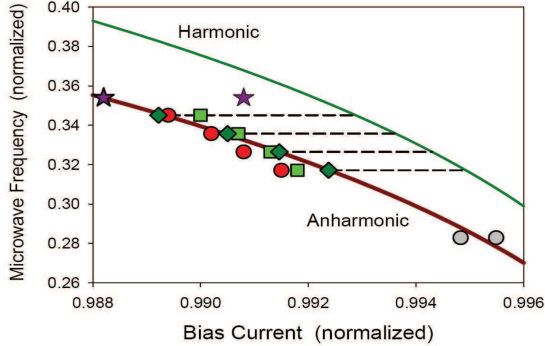}%
\caption{(color online) Comparison of classical model and quantum model with
experimental data. Circles are experimental points, squares are predictions of
the quantum model (digitized from the positions of the arrow markers in
Fig.3(b) in\cite{MDC}), and diamonds are predictions of the classical model
using the anharmonic approximation for resonance frequencies
Eq.(\ref{anharmonic}). \ The harmonic approximation, Eq.(\ref{Harmonic}) is
also plotted. \ The bottom pair of points are taken from the experimental data
in Fig.2 of \cite{MDC}; the upper pair of points (stars) are the two observed
peaks from Fig. 3 in \cite{Berkley}.}%
\label{Fig6}%
\end{center}
\end{figure}

In Fig.2, of \cite{MDC}, one of the experimental samples exhibited an
additional microwave escape peak. For that experiment, the sample parameters
were: $I_{C}=30.572\mu A$, $C=47pF$ and so $f_{J}=7.07GHz$. Hence the
normalized microwave frequency of $2GHz$ was $0.2829$. The two peaks were at
$30.43424\mu A$ and $30.41391\mu A$, which in normalized units are $0.995490$
and $0.994829$. These two points are included in Fig.\ref{Fig6}; they are both
quite close to the anharmonic curve.

In Fig.3 of \cite{Berkley}, there are two microwave induced escape peaks.The
sample parameters were: $I_{C}=14.12\ \mu A$, $C=4.2\ pF$ yielding a junction
plasma frequency $f_{J}=16.1\ GHz$. The microwave frequency was $5.7\ GHz$
which is $0.354$ in normalized units. The two peaks are at bias currents of
$I=13.9907\mu A$ and $I=13.9530\mu A$, which in normalized units are $0.9908$
and $0.9882$. \ These two points are included in Fig.\ref{Fig6}.
\ Interestingly, one appears to be on the anharmonic curve, while the other is
close to the harmonic curve, perhaps suggesting that the system can resonantly
respond to either condition. \ The possibility for multiple states in the ac
driven anharmonic potential is consistent with previously reported
observations (see Fig.1 in Ref.\cite{NGJ}). It should be pointed out that the
solid line Berkley's Fig.3 was described as \textquotedblleft a Lorentzian fit
to two peaks\textquotedblright, meaning it is not in any sense a theoretical
prediction and so does not constitute confirmation of quantum expectations.
The labelling of the two peaks as $\left\vert 1\right\rangle \rightarrow
\left\vert 2\right\rangle $ and $\left\vert 0\right\rangle \rightarrow
\left\vert 1\right\rangle $ is based on assumptions regarding applicable physics.

\

In the previous section, it was noted that the four single peaks, each at a
different microwave frequency, from Fig.3 of \cite{MDC} are reasonably
reproduced by classical calculations of relative escape rates. These classical
points are shown in Fig.\ref{Fig6} as diamonds that lie along the anharmonic
curve. \ However, the comparison of experiment with classical and quantum
theories described in \cite{MDC} contained the statement: \textquotedblleft
Furthermore, the measured positions of the resonances are clearly very
different from a classical prediction for the resonant activation of the
particleoscillating at the plasma frequency (dashed line)\textquotedblright.
\ This dashed line appears in part (b) of the figure and is in fact the
harmonic approximation. \ But we see from our Fig.\ref{Fig6} that the fair
test of the classical model is the anharmonic approximation, and at the very
least the classical model is as successful as the quantum hypothesis.

\section{Effect of Sweep Frequency}

It is important to note that peaks in the escape probability distributions are
not like lines in atomic spectra in that they are a manifestation of both the
fundamental physics associated with escape rates and the way the experiment is
performed - specifically, the frequency at which the bias current is swept
from zero to its critical value, ($f_{S}$).
\begin{figure}
[ptb]
\begin{center}
\includegraphics[
natheight=8.5in,
natwidth=11.0in,
height=2.6377in,
width=3.3935in
]%
{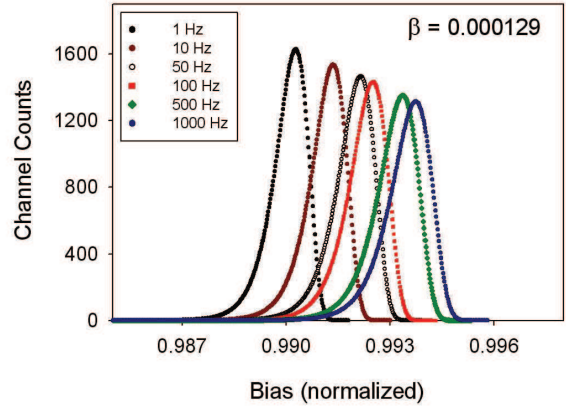}%
\caption{(color online) Simulation results showing the effects of bias sweep
frequency $f_{S}$ on the position of the peak in the escape probability
distribution.}%
\label{Fig7}%
\end{center}
\end{figure}
This issue was addressed in a swept bias simulation with: $N=50,000$,
$M=100,000$, $f_{J}=7.072\ GHz$, $f_{S}=10\ Hz$ and $\beta=0.00129$. The
results for microwaves OFF are shown in Fig.\ref{Fig7}. Clearly, the exact
location and shape of the thermal escape peak are determined in part by the
speed with which the bias current is ramped towards the critical value. As
might be anticipated, increasing the sweep rate moves the peak towards higher
bias currents.

\section{Effect of Dissipation}

To illuminate some aspects of the effects of dissipation, we carried out swept
bias simulations using Eqs. (\ref{GammaBHL}) and (\ref{at}), and for this
example with parameters $I_{C}=14.12\mu A$, $C=4.2pF$, $\beta=0.00018$. \ The
microwave frequency for the was set at $5.7GHz$ and the assumed excitation
parameters were $a=0.085$, $b=0.0012$. Typical values of the dissipation $Q$
for underdamped Josephson junctions lie in the range $20$ to $50$. Escape
histograms were repeated for a number of choices of the dissipation constant
parameter and the results are shown in Fig.\ref{Fig8}.
\begin{figure}
[ptb]
\begin{center}
\includegraphics[
natheight=8.5in,
natwidth=11.0in,
height=2.5356in,
width=3.1929in
]%
{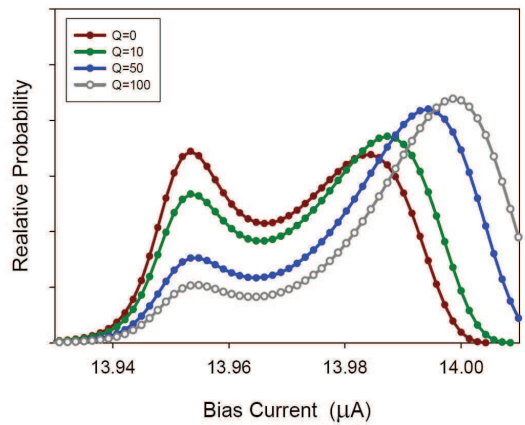}%
\caption{(color online) Swept bias simulations with selected values of the
parameter $Q$ using the escape rate due to B\"{u}ttiker, Harris, and
Landauer\cite{BHL}.}%
\label{Fig8.ps}%
\end{center}
\end{figure}
Note that the microwave peak does not shift, but the position of the purely
thermal peak varies with $Q$. Therefore, the particular value of the junction
dissipation might have a slight effect on predictions of the Voss \& Webb type
of experiment, but would not influence the positions of microwave induced peaks.

\section{Discussion}

The decades old papers of Voss and Webb, and Martinis \textit{et
al.}\ appeared to convincingly demonstate the (anticipated) appearance of MQT
in superconducting circuits operating below a crossover temperature. The
classical model was subsequently discarded as a possible source for observed
phenomena at millikelvin temperatures.\ In this paper we have shown that this
assertion may not be justified, and that these early foundational experiments
can certainly be modeled successfully within a purely classical device description.

With respect to the experiments of Voss and Webb, we have demonstrated that
there was not strong evidence that the junction had entered a macroscopic
quantum state even at the lowest temperatures. A classical model with some
self-heating gives a more consistent description of those observations.

With respect to the experiments of Martinis \textit{et al}., we have
demonstrated that in the presence of microwave irradiation the additional
peaks which appear in the swept bias escape distributions are just as well
accounted for within the classical resonant activation model as by the
proposed macroscopic quantum model.

The key issue in this situation was nicely expressed by Devoret, Martinis, and
Clarke \cite{DMC} as follows: \textquotedblleft An experiment cannot prove a
theory, but only invalidate an alternative theory.\textquotedblright. The
present study should therefore be seen in this context - the classical theory
for these systems has not yet been ruled out. Therefore, an exclusive
presumption of MQT in these systems is not justified.

\begin{acknowledgments}
We thank R.C. Ramos for providing the copy of his experimental data included
in Fig.5. This work was supported (JAB) by a grant from the Natural Sciences
and Engineering Research Council of Canada, and (MC) by a MIUR-PRIN08 program
(Italy). NGJ is grateful for support from Danmarks Nationalbank (Denmark).
\end{acknowledgments}

\end{document}